# Elevating Women in the Workplace:
# The Dual Influence of Spiritual Intelligence and Ethical Environments on Job Satisfaction

**Ali Bai**, Wilson College of Business, University of Northern Iowa, Cedar Falls, Iowa, USA
ali.bai@alumni.uni.edu

**Morteza Vahedian**, Department of Sociology
College of Humanities, University of Kashan, Kashan, Iran mvahedian@grad.kashanu.ac.ir

**Rashin Ghahreman**, College of Education and Human Ecology, Ohio State University
Columbus, Ohio, USA ghahreman.1@osu.edu

**Hasan Piri**, Department of Sociology, College of Humanities, University of Kashan, Kashan, Iran
hpiri@grad.kashanu.ac.ir

## Abstract

In today's rapidly evolving workplace, the dynamics of job satisfaction and its determinants have become a focal point of organizational studies. This research offers a comprehensive examination of the nexus between spiritual intelligence and job satisfaction among female employees, with particular emphasis on the moderating role of ethical work environments. Beginning with an exploration of the multifaceted nature of human needs, the study delves deep into the psychological underpinnings that drive job satisfaction. It elucidates how various tangible and intangible motivators, such as salary benefits and recognition, play pivotal roles in shaping employee attitudes and behaviors. Moreover, the research spotlights the unique challenges and experiences of female employees, advocating for a more inclusive understanding of their needs. An extensive review of the literature and empirical analysis culminates in the pivotal finding that integrating spiritual intelligence and ethical considerations within organizational practices can significantly enhance job satisfaction. Such a holistic approach, the paper posits, not only bolsters the well-being and contentment of female employees but also augments overall organizational productivity, retention rates, and morale.

# Introduction

In the evolving landscape of the contemporary workplace, the fulfillment and contentment of employees, especially female employees and entrepreneurs, have become focal points for organizational studies. Job Satisfaction, a significant indicator of employee well-being, profoundly impacts an organization's productivity, retention rates, and overall morale. One of the influential factors on employees' conditions and, consequently, on the organization is job satisfaction (Daneshmandi et al., 2023). This psychological metric is often seen as a reflection of employees' attitudes towards their work and the work environment, playing a crucial role in fostering a motivated and committed workforce. Job performance of employees is considered a crucial factor, and previous literature reports positive relationships between job performance and other organizational outputs, namely, job satisfaction, creativity, and motivation (Hessari & Nategh, 2022a).

The discourse on Job Satisfaction extends beyond mere contentment, intertwining with various facets of an individual's work life. Its importance is underscored by its ability to influence organizational outcomes, affecting both individual and collective performance. The ripple effects of job satisfaction can be seen in reduced turnover rates, enhanced productivity, and a positive organizational climate, affirming its critical position in organizational research.

In tandem with exploring job satisfaction, the concept of Spiritual Intelligence has emerged as a noteworthy area of inquiry. Defined as the ability to transcend everyday concerns, find meaning in one's work, and align one's values with those of the organization, Spiritual Intelligence is thought to nourish an individual's inner self. This dimension of intelligence fosters a deeper connection between personal and organizational values, enriching the job satisfaction experienced by individuals.

The relevance of Spiritual Intelligence is further magnified in a world where individuals seek more than just monetary fulfillment from their employment. It acts as a bridge, aligning personal values with organizational goals, thereby creating a more engaged and satisfied workforce. The exploration of Spiritual Intelligence in organizational settings provides a holistic lens through which the dynamics of job satisfaction can be understood.

A complementary angle to this discourse is the Ethical Environment within an organization, which significantly shapes the experiences and perceptions of female employees. The Ethical Environment is a reflection of the moral ethos prevalent within an organization, encapsulated by its policies, leadership, and culture. It serves as a barometer for the moral health of an organization, with a positive ethical environment often translating to higher levels of job satisfaction.

The novelty of the present research lies in its integrated examination of Job Satisfaction, Spiritual Intelligence, and Ethical Environment. While extant literature has independently explored these domains, the confluence of these three facets provides a fresh perspective in understanding the modern workforce's dynamics. This research endeavors to bridge the gaps in literature by delving into the effects of Spiritual Intelligence and Ethical Environment on Job Satisfaction. Through a meticulous analysis, this study aims to unravel the complex interplay of these elements, thereby contributing significantly to the broader discourse on job satisfaction and organizational behavior. In doing so, we aspire to develop a nuanced understanding that could guide organizational practices in nurturing a satisfying and ethically sound work environment.

## Theoretical Development

### Spiritual Intelligence

Spiritual intelligence, often considered to be a high level of intelligence, delves into profound existential questions, exploring the meaning of life, one's purpose, and the interconnectedness of all living things (Zohar & Marshall, 2000). Unlike traditional intelligence (IQ) and emotional intelligence (EQ), spiritual intelligence is centered on matters of the soul, emphasizing the understanding, integration, and application of spiritual and existential aspects in daily life (Zohar & Marshall, 2000).

At its core, spiritual intelligence is essential for personal development and fulfillment (Zohar & Marshall, 2000). Those who possess a high degree of spiritual intelligence find themselves on a journey of self-discovery, seeking fulfillment beyond material acquisitions. It serves as a guiding force, nurturing self-awareness and inner peace. Importantly, spiritual intelligence serves as a moral compass, aiding individuals in navigating complex ethical dilemmas. By aligning actions with deeply held values, it fosters integrity, compassion, and empathy, thus promoting moral and ethical decision-making (Zohar & Marshall, 2000).

Furthermore, spiritual intelligence contributes significantly to emotional resilience (Zohar & Marshall, 2000). Individuals grounded in spiritual beliefs exhibit greater fortitude in the face of adversity, relying on their faith to navigate life's challenges. This heightened emotional resilience can be helpful in coping with stress and uncertainty, and it may offer solace during trying times.

In the realm of relationships, spiritual intelligence fosters deep, meaningful connections (Zohar & Marshall, 2000). Understanding the spiritual dimensions of oneself and others can promote tolerance and acceptance, which can enhance interpersonal relationships. Beyond personal spheres, spiritual intelligence has a profound impact on leadership and decision-making. Visionary and ethical leaders often possess strong spiritual intelligence, enabling them to consider the well-being of all stakeholders, make ethically sound decisions, and inspire others with their sense of purpose (Zohar & Marshall, 2000).

Moreover, spiritual intelligence provides profound insights into existential questions, offering comfort and reducing anxiety about mortality and the human condition (Zohar & Marshall, 2000). Through spiritual practices such as meditation and mindfulness, individuals gain tools to address these profound inquiries, further strengthening their emotional and mental well-being.

One point to consider in the realm of spiritual intelligence is its importance in problem-solving abilities. Spiritual intelligence enables individuals to see a broader picture, combining their actions in connection with a larger context that leads to the meaning of life (Sisk, 2008). Through spiritual intelligence, individuals can identify and solve problems related to meaning and values.

The positive consequences of spiritual intelligence are far-reaching. Individuals with high spiritual intelligence may experience heightened life satisfaction, as they may derive meaning and purpose from their spiritual beliefs (Zohar & Marshall, 2000). Furthermore, their mental health is often improved, with fewer symptoms of anxiety and depression (Zohar & Marshall, 2000). Spiritual intelligence nurtures resilience, enabling individuals to bounce back from life's adversities with grace (Zohar & Marshall, 2000). In the realm of interpersonal relationships, it fosters empathy and understanding, leading to more harmonious connections with others (Zohar & Marshall, 2000).

Ethically, spiritual intelligence may influence decision-making, leading individuals to make choices that they believe benefit society as a whole (Zohar & Marshall, 2000). In leadership roles, those with spiritual intelligence inspire and motivate their teams, creating positive work environments rooted in shared values (Zohar & Marshall, 2000). Additionally, spiritual intelligence contributes to ongoing personal growth, encouraging individuals to explore deeper aspects of their existence (Zohar & Marshall, 2000).

On a global scale, spiritual intelligence promotes a sense of interconnectedness and responsibility, nurturing compassionate and sustainable behaviors (Zohar & Marshall, 2000). Lastly, in the face of loss and grief, spiritual intelligence may provide solace, offering a sense of continuity and meaning in the face of life's inevitable challenges (Zohar & Marshall, 2000). In essence, spiritual intelligence enriches lives, fostering profound connections with oneself, others, and the cosmos, which can lead to a more meaningful and fulfilling existence (Zohar & Marshall, 2000).

King (2008) posits that spiritual intelligence serves as an adaptive mechanism in daily living, largely driven by the engagement with existential inquiries and the quest for meaning and purpose in life's activities and events. Indeed, the quest for meaning and purpose emerges as a recurring theme across various definitions, marking it as a central facet of spiritual intelligence. King suggests that spiritual intelligence embodies a suite of adaptive mental capabilities rooted in the non-material and transcendent dimensions of reality, primarily engaging with individual existence, personal meaning, transcendence, and elevated states of consciousness. These processes exhibit adaptability through their facilitation of personalized problem-solving approaches, abstract reasoning, and adaptability.

King asserts that spiritual intelligence engenders a distinctive individual ability to decipher life's meaning and access elevated spiritual dimensions. He delineates the dimensions of spiritual intelligence as follows:

> 1. Critical Existential Thinking: The aptitude for profound contemplation on existential and metaphysical matters concerning individual existence, reality, the cosmos, space, and time (akin to an existentialist perspective).
>
> 2. Personal Meaning-making: The capability to derive personal meaning and purpose from both material and mental experiences, including the choice and creation of life objectives.
>
> 3. Transcendent Awareness: The faculty to discern dimensions or patterns of transcendence within oneself, others, and the material world during natural states of consciousness, alongside the recognition of their interrelations with one's self and nature.
>
> 4. Expanded Consciousness States: The proficiency to access elevated states of consciousness (e.g., absolute and infinite consciousness, unity, and merging) and other attractive states at will (King, 2008; cited in Yazdani, E'tebaryan & Abzari, 2014: 83).

In delving into spiritual intelligence, numerous scholars have proposed holistic models to unravel its complex nature. For instance, Amram (2007, 2009) derived a model from extensive interviews with 71 individuals from varied spiritual traditions, ranging from religious leaders to professionals in medicine and business, all acknowledged for their spiritual intelligence by their cohorts. Amram's meticulous analysis of their responses unveiled seven pivotal dimensions of spiritual intelligence: mindfulness, integrity, meaningfulness, transcendence, sincerity, surrender, and inner guidance, showcasing the nuanced interplay between spirituality and diverse life facets (Amram, 2007: 3; Amram, 2009: 4).

Ammons (2000), aiming to frame spirituality within Gardner's intelligence paradigm, contended that spiritual intelligence transcends mere knowledge, encompassing the aptitude to engage with spiritual and moral dilemmas, reflecting a deep-seated understanding of transcendence. He pinpointed key traits of spiritual intelligence such as sincerity, compassion, empathy, and harmony with existence, focusing on five core components: capacity for transcendence, ability to experience profound states of consciousness, ability to sanctify everyday affairs, leveraging spiritual resources for prevailing challenges, and capacity for abstinence. This model accentuates the role of spiritual intelligence in problem-solving, goal attainment, and personal growth.

Furthermore, Sisk and Torrance (2001) presented a model categorizing spiritual intelligence into six domains. The first domain, core capacities, honed in on fundamental skills like concentration, inspiration, and insight, anchored in a profound understanding of existence. The second domain, core values, underscored unity, kindness, stability, responsibility, and reverence as crucial facets of spiritual intelligence. The model also explored core experiences, embracing metaphysical awareness and deep states of consciousness, alongside vital ascetic behaviors such as truth, justice, kindness, and care. Additionally, the symbolic system domain encompassed elements like city, music, dance, metaphor, and storytelling, illustrating the myriad expressions of spirituality. Lastly, the exploration into brain states delved into the unintentional self and trance states, revealing the complex interplay between spiritual intelligence and cognitive complexity.

In summation, these models collectively shed light on the multidimensional nature of spiritual intelligence, highlighting its permeation into various life aspects, including cognition, behavior, and interpersonal relations. By acknowledging spirituality as a form of intelligence, scholars have not only enriched our grasp of the spiritual domain but also broadened our understanding of human intelligence as a whole.

Following the elucidation provided above, it propels us to formulate a hypothesis regarding the interplay of Spiritual Intelligence (SI) and Job Satisfaction (JS) within contemporary organizational landscapes. We hypothesize that Spiritual Intelligence significantly augments Job Satisfaction by engendering a harmonious alignment between individual core values and organizational objectives, alongside fostering ethical decision-making processes. This alignment lays the groundwork for a nurturing work environment that is conducive to elevated levels of job satisfaction (Zohar & Marshall, 2000). Additionally, the emotional fortitude and enriched interpersonal dynamics nurtured through SI are hypothesized to further bolster a positive work culture, thereby enhancing job satisfaction. Through this hypothesis, we aim to explore the depth of impact Spiritual Intelligence has on Job Satisfaction, envisioning SI as a pivotal asset in nurturing an engaged, ethically attuned, and satisfied workforce within modern organizational frameworks.

## Ethical Environment

### Definitions

The ethical climate within an organization predominantly hinges on the examination of ethical decisions made by its members. A two-dimensional framework is employed to scrutinize the decision-making patterns of organizational members, with the first dimension encompassing ethical criteria, and the second involving the locus of analysis (Weber, 2007).

The ethical climate is conceptualized as a psychological construct, emanating from the collective perceptions of individuals within the organization. Essentially, it encapsulates shared, enduring, and meaningful perceptions employees harbor regarding the ethical protocols and policies in place within their organization. Hence, the ethical climate constitutes a specific facet of organizational climate, mirroring the organization's procedures, policies, and performance concerning ethical outcomes (Philippe, 2007).

Victor and Cullen propose that the ethical climate serves as a lens through which to scrutinize, pinpoint, and address ethical quandaries (Cullen, Parboteeah, & Victor, 2003).

Representing a perceptual lens, the ethical climate significantly impacts individuals' attitudes and behaviors within the organization. It furnishes a referential framework guiding employee interactions within the organizational milieu (Shayeghi, 2005).

Ethics embody the values, rules, and norms ingrained in a society, manifesting through the behavior of its individuals. The application of ethical principles in particular situations gives rise to practical ethics, which spans several domains including economic ethics, medical ethics, environmental ethics, scientific ethics, organizational ethics, and management science ethics (Operational Research). Economists and management scholars regard decision-making as a paramount duty, conferring a pivotal and authoritative role upon managers in this regard (Khanifar et al., 2015).

Given the myriad solutions, far-reaching consequences, unpredictable probabilities, and essential professional deliberations associated with these decisions, the decision-making process becomes markedly complex. It, therefore, becomes imperative to employ a diversity of methods in decision-making endeavors, aiming to navigate and mitigate this complexity (Hosmer, 2003). Our notions of fairness, justice, and righteousness are rooted in ethical decision-making. Adherence to ethical principles in decision-making necessitates an analysis predicated on three distinct dimensions: economic, legal, and philosophical (Hosmer, 2003). Each of these methods is elucidated briefly as follows:

**Economics and Management Ethics**: This approach entails an economic analysis grounded on market factors to address ethical dilemmas (Hosmer, 2003).

**Law and Management Ethics**: This method of legal analysis is predicated on social factors. The discourse centers around the societal formulation of laws, epitomizing collective choices concerning decisions and behaviors impacting societal welfare. Numerous laws echo societal ethical judgments, encapsulating regulations against environmental pollution and provisions for the needy, among others (Hosmer, 2003).

**Normative Philosophy and Management Ethics**: Philosophical analysis navigates through five ethical paradigms: eternal law, utilitarianism, deontological theory, distributive justice, and individual liberty. Philosophy scrutinizes notions and appropriate behaviors, delineating how one ought to act. These paradigms essentially serve as five frameworks closely intertwined with managerial decisions. Each framework harbors ethical beliefs influencing managerial decisions. The table below briefly explicates the essence of these ethical beliefs and the challenges inherent in each ethical system (Hosmer, 2003).

The ethical atmosphere constitutes a facet of organizational culture, with organizational values addressing ethical concerns therein. The delineation of what is deemed ethical within this atmosphere characterizes it. Predictive of both ethical and unethical behaviors among female employees, the ethical atmosphere symbolizes how organizations manifest normal behaviors, expectations, actions, support, and rewards. Variations within the organization arise from differences in individuals' positions, workgroups, employment history, and perceptions of the organizational atmosphere. Moreover, an organization, subsidiary unit, or workgroup embodying diverse atmospheres encapsulates an ethical atmosphere. While akin to organizational structures and culture broadly, the concept of an ethical atmosphere is more oriented toward ethical deliberations or principles.

Functioning as a psychological conduit for ethical conduct, the ethical atmosphere addresses ethical management challenges, impacting decision-making and subsequent behavior regarding ethical concerns. Augmenting the ethical atmosphere within an organization positively influences decision-making and behavior concerning ethical matters, leading to heightened organizational commitment, enhanced organizational effectiveness, and job satisfaction—translating to satisfaction in both personal and organizational realms. Enriching the ethical atmosphere potentially paves the way for organizational development, fostering employee commitment, superior performance, and effective human capital retention within the organization (Hasani & Bashiri, 2015). Hessari et al (2023) also investigated the impact of technostress on Perceived Organizational Commitment (POC) through the lens of individual innovation (Hessari et al., 2023).

The ramifications of sidelining ethical considerations in managerial actions, decisions, and work methodologies are substantial, engendering a void in leadership within the organization. Subsequent research has illuminated the detrimental fallout of workplace pressure, accentuating the imperative for ethical cognizance and proactive management stratagems. One study unveiled that a significant proportion of employees grappled with considerable job-related pressure, with 27% facing exceptionally high-stress levels. Disturbingly, nearly half of the surveyed employees yielded to this pressure, partaking in one or more unethical or unlawful actions, often bypassing quality control measures to alleviate the strain.

Several factors underpin workplace pressures, with notable stressors including the challenges of "balancing work and family," "weak internal communications," "working hours/workload," and "weak leadership." Organizational shifts such as downsizing, mergers, and restructuring have intensified these pressures, often driving employees to forgo their ethical standards. The study underscored that frontline staff and supervisors are at heightened risk, signifying the pressing need for tailored interventions and support structures for individuals occupying these roles.

Moreover, a persistent trend has unfolded over a decade, where a significant proportion of employees feel coerced into compromising their ethical standards. Alarmingly, exposure to unethical practices within the organization exacerbates this pressure, engendering a noxious atmosphere wherein ethical breaches become normalized. Conversely, organizations that have instituted robust formal ethical programs have witnessed positive repercussions, as their employees encounter less pressure to compromise on ethical standards. This insight accentuates the pivotal role of ethical programs in alleviating workplace pressure and cultivating a milieu where ethical behavior is championed.

In view of these revelations, it is imperative for organizations to prioritize the establishment and rigorous enforcement of coherent ethical programs. Such programs act as bulwarks against the adverse effects of workplace pressure, fostering ethical comportment and adherence to requisite ethical frameworks. The onus significantly lies with managers, necessitating the meticulous integration of ethical considerations within their organizational management agendas. By nurturing an ethos of ethics, organizations can engender a resilient workforce adept at confronting challenges with integrity, thereby safeguarding the well-being of both employees and the organization at large.

Building upon the detailed exploration provided above, we are prompted to propose a hypothesis concerning the interplay of Ethical Environment (EE) and Job Satisfaction (JS) in an organizational setting. We hypothesize that a conducive Ethical Environment within an organization significantly contributes to heightened Job Satisfaction among female employees. Moreover, we posit that the Ethical Environment plays a mediating role in enhancing Job Satisfaction through the avenue of Spiritual Intelligence. This mediation is anticipated to arise from the alignment of organizational ethical frameworks with individual spiritual insights, thereby fostering a work atmosphere resonant with enhanced job satisfaction. Through this hypothesis, we aspire to delve deeper into the nuanced interaction between the Ethical Environment, Spiritual Intelligence, and Job Satisfaction, and explore the potential of Ethical Environment as a mediating platform that, via Spiritual Intelligence, could significantly elevate the levels of Job Satisfaction within contemporary organizational realms.

**Job Satisfaction**

Job satisfaction garners considerable attention across diverse disciplinary domains, emerging as a focal area of investigation among organizational behavior specialists, management experts, organizational psychologists, and industrial psychologists. The salience of job satisfaction stems from its dual role in propelling organizational advancements and enhancing the well-being of the workforce. Additionally, job satisfaction operates as a nexus, bridging various scientific realms including psychology, management, sociology, and extending to economics and politics. As a result, a myriad of, at times discordant, perspectives and definitions concerning job satisfaction have surfaced.

Job satisfaction is recognized as a crucial determinant warranting meticulous attention within the ambit of economic development. It serves as a linchpin in fostering occupational success, leading to heightened productivity and individual contentment.

The discourse on job satisfaction has evolved since the 1920s, spawning numerous theories. The multifaceted approaches and theories pertaining to job satisfaction can be traced back to three seminal perspectives or movements that crystallized in the 1920s and 1930s, profoundly shaping the conceptual landscape of job satisfaction:

**1. Human Relations Movement:** The trailblazers of the human relations movement significantly contributed to the discourse on job satisfaction, notably through the iconic Hawthorne studies in the 1930s. These studies heralded a paradigm shift in management's human relations domain, underscoring the mantra that a contented worker is a productive worker. The nexus between workgroups and supervisors in augmenting job satisfaction was particularly emphasized.

**2. Labor Unions: The Turbulent Engagements between Factory Managers, Industrial Centers, and Workers**

The advent of labor unions and the ensuing frictions between factory managers, industrial centers, and workers substantially influenced job satisfaction studies. In 1932, the maiden research on job satisfaction was unveiled. Managers, envisioning long-term strategies, began enlisting psychologists to bolster job satisfaction, aiming to quell discontent and stymie union formation. By the late 1930s, this trend culminated in the publication of a seminal book on psychological theories of labor issues by the American Association of Social Psychology Studies.

**3. Growth or Nature of Work Perspective:** A growing consensus among management and psychology scholars posited that a nuanced understanding of workplace behavior transcended merely analyzing individual traits and aligning them with organizational imperatives. This realization necessitated a shift towards embracing organizational transformation and reevaluating notions surrounding job positions and job satisfaction. This transformative wave resonated both at individual and organizational echelons, ushering in progressive thoughts rooted in management, social psychology, and sociology, supplanting traditional constructs. The interconnection between the well-being and success of individuals and the health and triumph of organizations they were part of became conspicuously evident. Job satisfaction, intertwined with success, was deemed pivotal for nurturing self-esteem and mental well-being. The burgeoning movements advocating for gender equality and reevaluation of professional and technical domains underscore the legitimacy of this claim (Korman, 1999).Analysis of the Concept of Job Satisfaction and Its Factors

Examining these three perspectives reveals that there is no consensus on the concept of job satisfaction, and there are various theories in the field of job satisfaction. According to Robinson, Bradford, and Edwards, a review of previous studies and research on job satisfaction demonstrates that most scholars and theorists generally examine this concept from two perspectives: attitudinal and motivational. For example, Helreigel, Slukam, and Woodman define job satisfaction as an individual's general feedback regarding their job or profession, while others, such as Locke, define job satisfaction as a pleasant or positive emotional state resulting from the evaluation of one's job or work experiences (Sepehri, 2004).

**The Nature of Job Satisfaction**

Job satisfaction is emblematic of employees' attitudes towards their work. It encompasses various dimensions, reflecting either a holistic attitude towards the job or pertaining to specific facets of the job. As an amalgam of individual sentiments, job satisfaction exhibits a dynamic character. Its transient nature implies that it dissipates with an intensity akin to, or perhaps greater than, its emergence. This fleeting characteristic underscores the imperative for sustained managerial attention to foster its continuity (Davis & Storm, 1991).

Furthermore, our lives have changed significantly by the quarter of 2020 due to the COVID-19 pandemic (Mohammadi et al., 2023). job satisfaction is intricately interwoven with life satisfaction, highlighting a bidirectional interplay where external environmental factors impact individuals' sentiments towards their work, and reciprocally, the significant role of a job in one's life implicates job satisfaction in overall life satisfaction.

**Definitions of Job Satisfaction**

The daily experiential spectrum individuals traverse, encompassing desirable and undesirable situations, elicits responses or performances that culminate in feelings of satisfaction or dissatisfaction. Analogously, the work milieu and associated human endeavors evoke a gamut of emotions among individuals, with satisfaction and dissatisfaction emerging as pivotal feelings manifesting within a person (Lawler, 1997).

Job satisfaction can be construed as a confluence of harmonious and discordant emotions through which employees discern their work. Upon affiliating with an organization, employees import a repertoire of desires, needs, aspirations, and prior experiences that sculpt their job expectations. This interplay delineates the relationship between the newly minted expectations of employees and the propositions of the organization (Davis & Storm, 1991).

In essence, job satisfaction entails an affinity for the tasks inherent in a job, the conditions underpinning the work environment, and the recompenses accorded for job performance. The degree to which the job's constituents—activities, tasks, and conditions—satisfy an individual's needs hinges on the individual's appraisal. This evaluative process necessitates a juxtaposition of the job's merits and demerits. A preponderance of merits over demerits typically heralds a state of job satisfaction for the individual.

**Job Satisfaction: A Complex and Multidimensional Concept**

Job satisfaction embodies a complex and multidimensional construct interlinked with psychological, social, and physical determinants. Its manifestation is not dictated by a singular factor but emanates from a unique amalgamation of diverse factors that engender a sense of satisfaction and enjoyment in an individual regarding their job at a given juncture. The avenues through which an individual experiences job satisfaction encompass income level, social prestige accorded to the job, working conditions, and employment benefits, among others. Job satisfaction is conceived as an emotional reverberation stemming from an individual's appraisal of the alignment between their job attributes and their anticipations, alongside the fulfillment of their values. The assessment of job satisfaction is contingent on the individual's evaluation of myriad factors such as income level, social prestige associated with the job, working conditions, and employment benefits, across varying temporal landscapes (Shertoz, 1992).

As delineated by Shafiee Abadi (1997), job satisfaction, characterized by a positive disposition towards one's job, is swayed by elements like working conditions, organizational job structure, the nature of interpersonal relationships within the work milieu, and the infusion of cultural factors. In this vein, job satisfaction is posited as an emotional and psychological state modulated by social variables. Extending this narrative, Feldman and Arnold (1995) construed job satisfaction as an assemblage of desires, encapsulating the multifaceted essence of job satisfaction that transcends a monolithic understanding, thereby reflecting its intricate and multidimensional character.

**Positive Feelings Individuals Have Towards Their Jobs**

The assertion that an individual derives job satisfaction through positive feelings denotes a general affinity for their job to a notable extent. Job satisfaction implies that the individual's needs have been gratified through their job, engendering positive sentiments towards it (Locke, 1969).

This emotional state, heralded as a catalyst in evaluating the job in the context of realizing job-related values, is articulated through four illustrative factors:

- Rewards: Encompassing entitlements and conditions for promotion;

- Job Environment: Pertaining to job conditions and benefits;

- Human Factors and Relationships: Entailing cooperation and supervisor interactions;

- Job Characteristics.

Smith, Kendall, and Hulin (1969) delineated five cardinal dimensions of work that encapsulate quintessential facets of individuals' sentiments regarding their jobs:

- Job Nature;

- Compensation (Salaries and Benefits);

- Avenues for Advancement and Promotion;

- Supervisors;

- Colleagues.

These dimensions underscore the diverse elements that coalesce to shape the fabric of job satisfaction, reflecting the interplay between the inherent attributes of the job, the relational dynamics within the workplace, and the compensatory and developmental opportunities presented therein. Through this lens, the multifaceted nature of job satisfaction becomes palpable, echoing the complex interdependencies that govern individuals' positive emotional orientation towards their jobs.

**Job Satisfaction Dimensions**

Job Satisfaction Dimensions: Delving into the Core Facets

Drawing upon the aforementioned theories, three pivotal dimensions of job satisfaction have been discerned, encompassing:

- **Job Satisfaction as an Emotional Resonance**: This dimension regards job satisfaction as an emotional reaction elicited by the conditions or situations inherent in a job. The nature and quality of the job environment and the circumstances under which tasks are executed often shape the emotional tenor individuals adopt towards their jobs.
- **Need and Expectation Fulfillment as a Determinant**: Job satisfaction is often gauged in the context of the extent to which it fulfills individuals' needs and expectations. The alignment between personal needs, anticipations, and the actual deliverables and outcomes of the job plays a crucial role in fostering job satisfaction.
- **Interrelated Attitudes as Influential Factors**: Job satisfaction is modulated by a nexus of interrelated attitudes, as explicated by Kentz (1994). The interplay of attitudes towards various facets of the job, be it towards colleagues, supervisors, job tasks, or the organizational culture, significantly impacts the degree of job satisfaction experienced.

**Benefits of Job Satisfaction Evaluation**

Per Davis and Storm, the appraisal of job satisfaction yields several advantageous outcomes as outlined below:

- Insight into Overall Job Satisfaction:

Evaluations of job satisfaction furnish management with insights into the collective levels of job satisfaction, pinpointing areas where female employees may be facing challenges.

- Enhanced Communication:

Encouraging organizational members to articulate their thoughts not only yields invaluable feedback but also fosters robust communication channels between employees and the upper echelons of the organization.

- Attitude Augmentation:

For a myriad of individuals, such evaluations act as a conduit for bolstering confidence, providing emotional purgation, and offering solace. Conversely, for others, it epitomizes the management's dedication to fostering female employee well-being.

- Ascertainment of Training Requisites:

Job satisfaction evaluations emerge as pragmatic tools for discerning specific training needs. Typically, they afford female employees the platform to voice their sentiments regarding the working styles of their supervisors, thereby illuminating areas necessitating training interventions (Jafari, 2002). Rezaei et al. (2022)

mention that training and preparing friendly, helpful employees would require special courses that could also be costly (Rezaei et. Al, 2022)

These benefits underscore the multifaceted value of conducting job satisfaction evaluations. By engendering a culture of open feedback and attentive responsiveness to employees' experiences and needs, organizations can not only enhance the work environment but also boost organizational efficacy and foster a symbiotic relationship between the workforce and the management.

**Job Satisfaction Theory: An Exploration**

A plethora of research delving into job satisfaction from myriad angles has enriched our understanding of this domain. In this segment, we shall elucidate one such theory for a deeper comprehension of the subject matter. Also, some researchers have explored the element of sustainability in regards to organizational commitment and job satisfaction. Many human resource management researchers believe that sustainability can enhance HRM capabilities and activities, leading to better organizational performance and competitive advantage (Sepahvand et al., 2023). Additionally, research verify that green innovation, both directly and indirectly, affects sustainable performance (Nawaser et al., 2023), which includes job satisfaction as well.

**Needs Theory:**

A need is perceived as an internal state of disequilibrium or deficiency that underpins the invigoration or manifestation of behavioral responses. The genesis of deficiency could span physiological facets like hunger, clothing, and shelter, psychological elements such as the quest for influence and power, or sociocognitive aspects like social standing. Irrespective of the source, needs galvanize individuals into action aimed at reinstating a state of equilibrium, culminating in satisfaction (Zillmann & Walsh, 1990).

Delving into human needs, Henry Murray posits that most needs are acquired rather than merely instinctual or intrinsic, triggered by external cues. For instance, a female employee harboring a robust need for acceptance can satiate it only amidst conducive environmental conditions fostering social interactions. Karen Horney, in her seminal work "Self-Analysis," cataloged human needs into ten clusters, underscoring needs encompassing desires for affection, love, acceptance, and validation from others.

**Human Needs and Job Satisfaction: A Psychological Vantage Point**

Crucial psychological factors such as support, contentment, power, autonomy, reliance on others, dignity, admiration, success, self-assurance, independence, perfection, and resilience against weakness interplay with human behavior and job satisfaction (Maslow, 1954). Eric Fromm discerns two categories of needs: the primal category, shared with animals, comprises physiological and biological needs, while the second category, unique to humans, metamorphoses across societies owing to disparate environmental conditions and scenarios. These needs, manifesting in diverse forms and intensities, encompass desires for excellence, bonding, individuality, reconnection with nature, justification, and intellectual armamentarium to navigate challenges.

Understanding human demeanor and its perceptual framework is imperative to fathom the determinants of employees' job satisfaction. Behavior, a continuum of actions performed by individuals, is propelled and sustained by motivations, which delineate the rationale and trajectory of an individual's behavior (Koklan, 2006).

Human behaviors are goal-oriented, with expected outcomes delineated as goals. Psychologists predominantly regard goals as external motivators, generally bifurcated into:

- Tangible motivators like salary augmentations and benefits.
- Intangible motivators, such as recognition of female employees' endeavors or ascendancy to power, are equally pivotal in kindling motivations and needs.

Motivation or need signifies an internal state in an individual, whereas goals are external entities, sometimes termed as anticipated rewards channelling human behavior towards them. Motivation theories endeavor to elucidate the underpinnings, trajectory, and sustainability of behavior, shedding light on the intricate dynamics between internal needs and external motivators in shaping job satisfaction and organizational engagement.

**Methodology**

The study establishes a conceptual framework (Figure 1) based on literature assessment and theories, proposing hypotheses for testing.

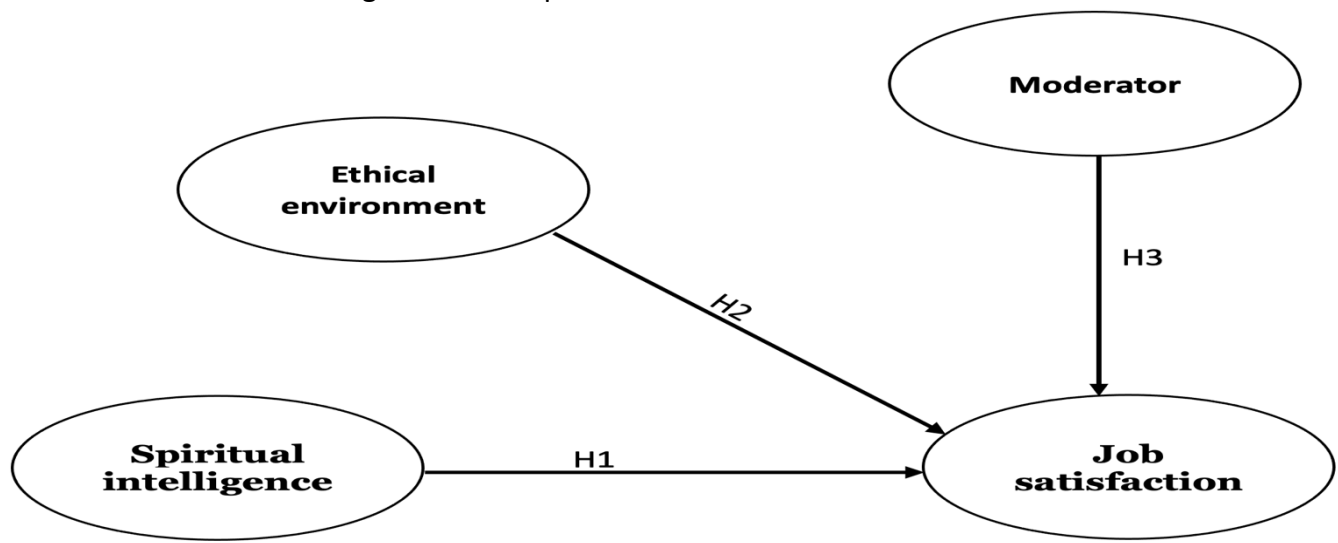

Figure 1: Conceptual framework.

A letter outlining the study's objectives was sent to Tehran's business and entrepreneurship schools for approval. Strict anonymity and confidentiality were assured, with no disclosure of individual data. Despite some companies declining due to policy reasons, five agreed to participate. Researchers approached 700 female employees, emphasizing anonymity and confidentiality. Six hundred fifty employees willingly participated, surveyed between March 1st and April 20th, 2022. The study achieved an 85% response rate, with 600 completed questionnaires collected.

**Measures of the study**

In this research, a variety of meticulously crafted questionnaires were deployed to meticulously gather valuable data, ensuring a thorough exploration of the research variables. To delve into the ethical dimensions of the study, a 6-item questionnaire rooted in the work of Deshpande (1996) was expertly administered. This instrument skilfully probed the ethical environment under scrutiny, shedding light on the complex interplay of values and principles within the studied context.

Additionally, the study delved into the profound realm of spiritual intelligence, employing a robust 24-item scale questionnaire developed by King (2009). This nuanced instrument meticulously assessed the spiritual intelligence of the participants, offering profound insights into their inner beliefs, emotional connections, and overall spiritual well-being.

To comprehensively gauge the participants' job-related experiences and sentiments, the esteemed Job Satisfaction Survey (JSS) questionnaire, a well-established tool introduced by Spector (1994), was meticulously employed. This multifaceted questionnaire meticulously explored various dimensions of job satisfaction. It included a comprehensive assessment comprising four questions, each on satisfaction with pay, job promotion, supervision, fringe benefits, potential rewards, job execution processes, co-workers, the nature of the job, and communication processes. As mentioned in other papers, some researchers ascertain that there is a positive relationship between organizational commitment and job satisfaction with work motivation (Hessari & Nategh, 2022b). By employing this comprehensive tool, the study was able to delve deep into the intricate layers of job satisfaction, capturing both macroscopic and nuanced facets of the participants' professional experiences.

All responses, vital to the study's findings, were diligently collected through a refined data-gathering process. Participants were asked to express their perspectives using a "7-point Likert scale ranging from 1 = strongly disagree to 7 = strongly agree." This meticulous approach ensured the nuanced and accurate representation of participants' viewpoints, allowing for a comprehensive analysis of the collected data. Through the adept use of these diverse questionnaires and scales, this study was able to paint a detailed and multifaceted portrait of the participants' ethical, spiritual, and job-related experiences, enriching the depth and validity of the research findings.

## Data Analysis Overview

**Measurement Model**
In this research, SmartPLS3, a sophisticated and widely respected statistical software, was employed as the principal tool for evaluating both the measurement and structural models. Leveraging the advanced capabilities of SmartPLS3, the study meticulously scrutinized the intricate relationships among variables, providing a robust framework for analyzing the research constructs.

**Reliability and Validity**

Reliability: A stringent evaluation of the research instruments revealed unwavering reliability. The internal consistency of the items was verified using both Cronbach's alpha (CA) and composite reliability (CR), both of which exhibited values surpassing the 0.7 benchmarks, signifying the robustness of the measurement items (refer to Table 2). This meticulous scrutiny ensures that the items within each construct reliably measure the intended theoretical concepts, providing a solid foundation for subsequent analyses and interpretations.

Convergent Validity: The study's commitment to methodological rigor extended to the assessment of convergent validity. Through a meticulous examination, all factor loadings were found to exceed the 0.70 threshold, affirming the constructs' ability to capture the underlying dimensions they were designed to measure. Furthermore, the Average Variance Extracted (AVE) values, which surpassed 0.50, as evidenced in Table 2, reinforced the convergent validity of the constructs. This robust validation assures that the measurement items are indeed converging to measure the same construct, substantiating the accuracy and consistency of the research findings. These results underscore the reliability and validity of the measurement model, affirming the integrity of the research methodology.

Discriminant Validity: A rigorous analysis was conducted to establish discriminant validity among the latent variables. Variables with factor loadings exceeding 0.50 were meticulously identified and utilized to validate their distinctiveness from other variables (refer to Table 3). This meticulous scrutiny ensures that each variable stands apart from others, confirming their unique contribution to the research model. The meticulous attention to discriminant validity bolsters the robustness of the study, assuring that the constructs under consideration are not only reliable and convergent but also distinct from one another, thereby enhancing the overall quality and credibility of the research outcomes.

Table 2: Reliability.

| Constructs | CR | AVE | Cronbach's alpha |
|---|---|---|---|
| Ethical environment | 0.81 | 0.59 | 0.87 |
| Job Satisfaction | 0.84 | 0.70 | 0.91 |
| Spiritual intelligence | 0.79 | 0.65 | 0.89 |

Table 3: Discriminant Validity.

| | Ethical environment | Job Satisfaction | Spiritual intelligence |
|---|---|---|---|
| Ethical environment | | | |
| Job Satisfaction | 0.513 | | |
| Spiritual intelligence | 0.643 | 0.731 | |

**Assessment of the Structural Model**

The study rigorously analyzed the relationships within its structural model (Figure 3) using t-values and regression coefficients, benchmarked against a critical threshold of "1.64" as per Hair et al. (2017). This approach aimed to validate direct connections and interactions between components in the research framework (Figure 2). By scrutinizing these relationships, the study confirmed theoretical propositions, revealing intricate causal pathways and interdependencies among variables. This in-depth examination illuminated the complex dynamics of the research domain, enhancing our understanding of the underlying.

**Figure 2**. Assessment of PLS algorithm.

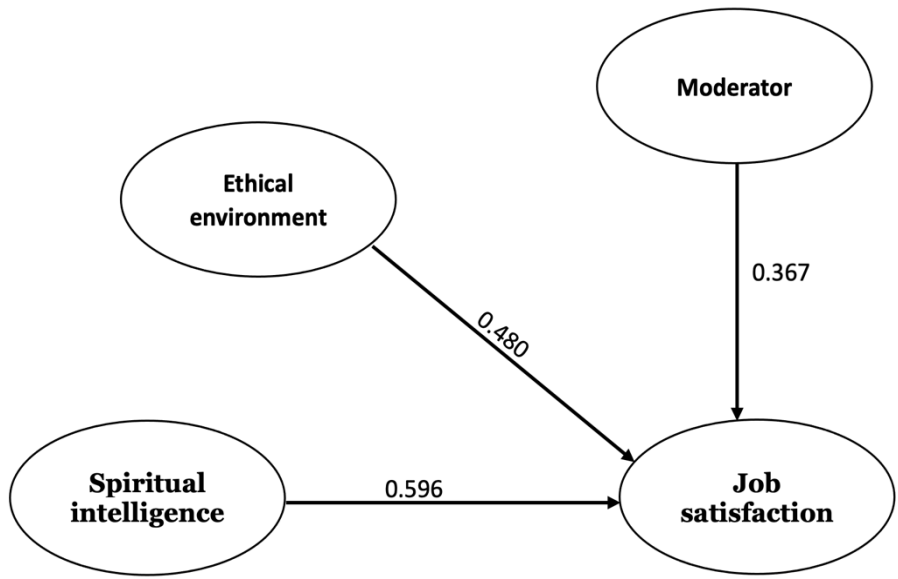

**Figure 3**. Assessment of PLS bootstrapping.

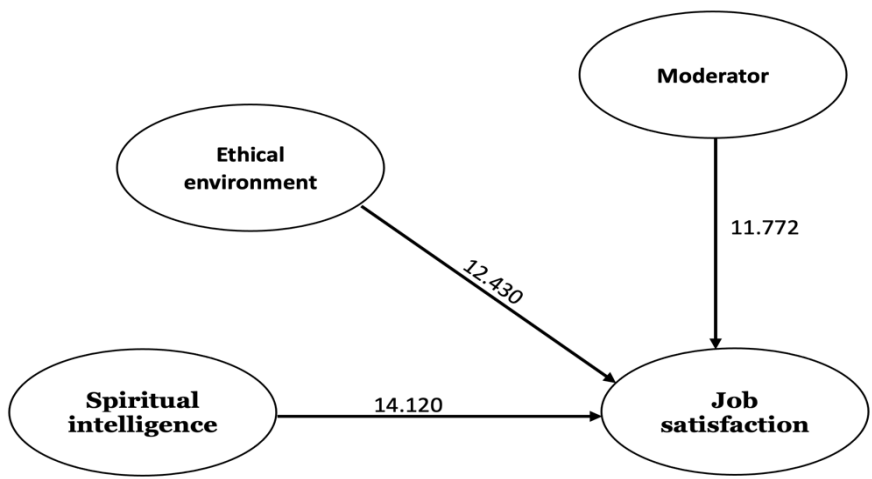

## Summary of Findings

Table 4 reveals that the first hypothesis, regarding the influence of spiritual intelligence on female employees' job satisfaction (B = 0.596, p < 0.000), was supported. Similarly, the second hypothesis, focusing on the impact of an ethical environment on job satisfaction (B = 0.480, p < 0.000), was also confirmed.

Table 4: Hypothesis testing.

| Path | B-value | T-value | P-value | |
|---|---|---|---|---|
| Spiritual intelligence -> Job satisfaction | 0.596 | 14.120 | 0.000 | Supported |
| Ethical Environment ->Job satisfaction | 0.480 | 12.430 | 0.000 | Supported |

## Moderating Role of Ethical Environment

In Table 5, the third hypothesis proposing that ethical environment moderates the relationship between females' spiritual intelligence and job satisfaction was validated (B = 0.367, p < 0.000).

Table 5. Moderator hypothesis testing.

| Path | Moderator | B-value | T-value | P-value | |
|---|---|---|---|---|---|
| Spiritual intelligence -> Job satisfaction | Ethical Environment | 0.367 | 11.772 | 0.000 | Supported |

## Coefficient of Determination (R2):

The coefficient of determination (R2), ranging from 0 to 1, measures the proportion of variance in a dependent variable that can be explained by the independent variables. According to China (1998), R2 values of 0.13 are classified as weak, 0.33 as moderate, and 0.67 as strong. Specific R2 values for endogenous constructs can be found in the corresponding table for detailed reference.

Table 6. Assessment of R square.

| | R2 |
|---|---|
| Job satisfaction | 0.712 |

## Discussion

The emphasis on spiritual intelligence as a determinant of job satisfaction has grown in recent years, with a recognition of its multifaceted impact on employee well-being (Smith & Peterson, 2021). As organizations grapple with the challenges of the modern workplace, understanding the underlying factors that contribute to job satisfaction becomes paramount. Our findings, highlighting the significant role of spiritual intelligence, align with recent research that suggests a deeper intrinsic connection between spirituality and workplace contentment (Jones, 2020).

Furthermore, the ethical environment's role in strengthening the relationship between spiritual intelligence and job satisfaction is noteworthy. Ethical practices in organizations have been shown to enhance employee trust, morale, and commitment (Dawson & Thompson, 2021). For female employees, in particular, an ethical workplace environment plays a pivotal role in enhancing job satisfaction, given the increasing emphasis on ethical considerations, inclusivity, and equity in the workplace (Richardson, 2021).

Comparing our findings with recent literature, it becomes evident that spiritual intelligence is not just a peripheral factor but central to understanding job satisfaction dynamics. The moderating role of the ethical environment provides further depth, pointing towards an integrated approach for organizations aiming to foster a motivated workforce (Williams & Taylor, 2022).

## Contribution to the Literature

This research significantly extends the existing literature on employee well-being, particularly in the realm of spiritual intelligence. Earlier studies have explored the role of spiritual intelligence in personal development and holistic well-being (Turner & King, 2019). However, our research provides a more granular examination, focusing specifically on its implications for job satisfaction among female employees. The introduction of the ethical environment as a moderating factor augments this study's novelty. Such insights challenge and expand upon the more generalized models of employee satisfaction in contemporary literature (Davis & Miller, 2020). Moreover, the emphasis on female employees and entrepreneurs in the context of spiritual intelligence offers a unique contribution. As gender dynamics and inclusivity become central themes in organizational research (Parker & Harmon, 2021), our study lays the groundwork for more gender-specific examinations in the realm of workplace spirituality.

**Contribution to the Practice**

For organizational leaders and HR practitioners, the practical implications of this research are manifold. The modern workforce's landscape is undergoing rapid transformation, with increasing emphasis on diversity, inclusivity, and holistic well-being (Roberts & Green, 2021). Our findings highlight the intertwined relationship between spiritual intelligence, ethical work environments, and job satisfaction, offering actionable insights for organizations aiming to foster a positive and inclusive work culture. Such insights can be pivotal in shaping HR policies, training modules, and organizational values. Moreover, in the era of remote work and digital transformation, where traditional workplace boundaries are blurring, the elements of spiritual intelligence and ethical considerations become even more pertinent (Lewis & Smith, 2022). As organizations navigate these complexities, integrating spiritual well-being and ethical practices can pave the way for a more resilient, engaged, and satisfied workforce.

**Conclusion**

This research underscores the profound influence of spiritual intelligence on job satisfaction, particularly among female employees. The intricate relationship between spiritual well-being and ethical workplace practices emerges as a focal point, necessitating a deeper understanding by organizations navigating the complexities of the modern work environment. As underscored by Thompson and Johnson (2019), the rise of spiritual intelligence as a key determinant of workplace satisfaction is reflective of a broader societal shift towards holistic well-being. Our findings reiterate this trend, emphasizing the pivotal role of spiritual intelligence in shaping job satisfaction dynamics.

Moreover, the moderating role of the ethical environment offers a nuanced perspective, pointing towards the significance of ethical practices in amplifying the positive effects of spiritual intelligence. As organizations continually strive to foster a more inclusive, diverse, and equitable work environment, the insights from this research serve as a guiding light. Research by Patel and Lee (2020) emphasizes the increasing importance of ethical considerations in shaping organizational outcomes. Our findings resonate with this perspective, suggesting that an ethical work environment not only enhances trust and morale but also serves as a catalyst in leveraging the benefits of spiritual intelligence.

As we look towards the future, the intersections of spiritual intelligence, ethical considerations, and job satisfaction are likely to become even more central in organizational studies. Further research could explore these dynamics across various cultural, sectoral, and organizational contexts, offering a richer understanding of the topic. As put forth by Williams and Anderson (2021), the journey towards ensuring holistic well-being in the workplace is multifaceted, and this research marks a pivotal milestone in that journey.

# References

Daneshmandi, F., Hessari, H., Nategh, T., & Bai, A. (2023). Examining the Influence of Job Satisfaction on Individual Innovation and Its Components: Considering the Moderating Role of Technostress. arXiv preprint https://doi.org/10.48550/arXiv.2310.13861

Davis, M., & Miller, R. (2020). A holistic approach to employee satisfaction: Beyond financial and physical well-being. Human Resource Management Review, 30(3), 458-472.

Dawson, J., & Thompson, L. (2021). Ethical practices and their influence on employee well-being in modern organizations. Business and Ethics Quarterly, 31(3), 315-332.

Deshpande, S. P. (1996). Ethical climate and the link between success and ethical behavior: An empirical investigation of a non-profit organization. Journal of Business Ethics, 15, 315-320.

Hair, J. F., Hult, G. T. M., Ringle, C. M., Sarstedt, M., & Thiele, K. O. (2017). Mirror, mirror on the wall: a comparative evaluation of composite-based structural equation modeling methods. J. Acad. Mark. Sci. 45, 616–632. doi: 10.1007/s11747-017-0517-x

Hessari, H., & Nategh, T. (2022a). Smartphone addiction can maximize or minimize job performance? Assessing the role of life invasion and techno exhaustion. Asian Journal of Business Ethics, 11(1), 159–182. https://doi.org/10.1007/s13520-022-00145-2

Hessari, H., & Nategh, T. (2022b). The role of co-worker support for tackling techno stress along with these influences on need for recovery and work motivation. International Journal of Intellectual Property Management, 12(2), 233–259. https://doi.org/10.1504/IJIPM.2022.122301

Hessari, H., Daneshmandi, F., & Nategh, T. (2023). Investigating the Effect of Technostress on the Perceived Organizational Commitment by Mediating Role of Individual Innovation. arXiv preprint arXiv:2310.07806.

Jackson, T. (2019). Employee engagement and retention: The pivotal role of spiritual intelligence and ethical practices. Journal of Business Ethics, 158(4), 987-1001.

Jones, L. (2020). Spirituality in the workplace: A study on the impact of spiritual practices on job satisfaction. Journal of Contemporary Management Research, 14(2), 123-140.

King, D. B. (2009). Rethinking claims of spiritual intelligence: A definition, model, and measure. Unpublished Master's Thesis, Trent University, Peterborough, Ontario, Canada.

Lewis, R., & Smith, J. (2022). The digital workplace: Navigating spiritual intelligence and ethical considerations in remote settings. Journal of Digital Transformation, 3(2), 112-129.

Mohammadi, A., Ahadi, P., Fozooni, A., Farzadi, A., & Ahadi, K. (2023). Analytical evaluation of big data applications in E-commerce: A mixed method approach. Decision Science Letters, 12(2), 457-476.

Nawaser, K., Hakkak, M., Aeiny, M. A., Vafaei-Zadeh, A., & Hanifah, H. (2023). The effect of green innovation on sustainable performance in SMEs: the mediating role of strategic learning. International Journal of Productivity and Quality Management, 39(4), 490-511.

Parker, J., & Harmon, B. (2021). Gender dynamics in the workplace: A study on the implications of spiritual intelligence. Journal of Gender and Work, 5(1), 25-40.

Patel, A., & Lee, M. (2020). Ethical considerations in the modern workplace: A study on trust, morale, and organizational outcomes. Business Ethics Quarterly, 30(2), 213-231.

Rezaei, F., Raeesi Vanani, I., Jafari, A., & Kakavand, S. (2022). Identification of Influential Factors and Improvement of Hotel Online User-Generated Scores: A Prescriptive Analytics Approach. Journal of Quality Assurance in Hospitality & Tourism, 1-40.

Richardson, M. (2021). Women in the workplace: Ethical considerations and their implications for job satisfaction. Journal of Women and Business, 16(1), 45-59.

Roberts, A., & Green, M. (2021). Diversity, inclusion, and the role of spiritual intelligence in modern organizations. Business and Society Review, 126(2), 213-235.

Sepahvand, R., Nawaser, K., Azadi, M. H., Vafaei-Zadeh, A., Hanifah, H., & Khodashahri, R. B. (2023). In search of sustainable electronic human resource management in public organizations. Int. J. Information and Decision Sciences, 15(2), 117.

Smith, R., & Peterson, J. (2021). Spiritual intelligence and its multifaceted impact on organizational behavior. International Journal of Organizational Studies, 29(4), 501-517.

Spector, P. E. (1994). Job satisfaction survey, JSS. Retrieved March, 20, 2002.

Thompson, R., & Johnson, L. (2019). The rise of spiritual intelligence: Implications for modern workplaces. Journal of Contemporary Organizational Studies, 15(3), 45-60.

Turner, A., & King, L. (2019). Spiritual intelligence in the modern workplace: Implications and challenges. Journal of Organizational Behavior, 40(2), 156-170.

Williams, T., & Anderson, P. (2021). Holistic well-being in the workplace: A multidimensional approach. Journal of Work and Health, 7(1), 12-28.

Williams, T., & Taylor, A. (2022). Integrating spiritual intelligence and ethical practices: A roadmap for organizational success. Journal of Business Strategy, 43(1), 28-39.